\documentstyle [12pt,fleqn,epsfig]{article}
\begin{document} 
\baselineskip 20pt



\title{Resonant interactions in B\'{e}nard-Marangoni 
convection in cylindrical containers}

\author{B. Echebarr\'{\i}a $\ast$, D. Krmpoti\'{c} $\star$ 
and C. P\'{e}rez-Garc\'{\i}a $\ast$ \\
$\ast$ \small Departamento de F\'{\i}sica y 
Matem\'{a}tica Aplicada, Facultad de Ciencias,\\ 
\small Universidad de Navarra, 31080 Pamplona, 
Navarra, Spain.\\
$\star$ \small Departamento de F\'{\i}sica, 
Universidad Nacional de La Plata,\\ 
\small La Plata, Argentina.}

\maketitle

\begin{abstract}
Convection in a cylindrical container of small aspect 
ratio is studied. It is known that when, in addition 
to buoyancy forces, thermocapillarity effects are 
taken into account, resonant interactions of two 
modes may appear. In the case of 1:2 resonance 
amplitude equations are derived, showing the 
existence of a stable heteroclinic 
orbit and rotating waves, until now not observed 
experimentally. 
\end{abstract}

\vskip1cm

\centerline{Keywords:  Convection, Hydrodynamic instability, bifurcations, 
nonlinearity.}

\vskip0.5cm

\centerline{P.A.C.S.: 44.25.+f, 47.20.Dr, 47.20.Tg} 

\newpage

\section{Introduction}

Since the works of B\'{e}nard \cite{Benard} 
at the beginning of this century, thermal 
convection became one of the most 
studied pattern forming systems \cite{Cross}. 
In B\'{e}nard-Marangoni (BM) 
experiments a fluid layer confined in a container open to the 
atmosphere is heated from below. When the 
difference of temperature between the top and the bottom 
surface (i.e. $\Delta T$) is bigger than a critical 
value convection appears. Pearson \cite{Pearson} showed that 
besides buoyancy forces, thermocapillarity effects provide another
instability mechanism in this case. 
	
For shallow layers buoyancy forces dominate,
while for thick layers thermocapillarity is 
the main mechanism. Two 
different regimes appear depending on the 
liquid depth and the aspect 
ratio $a$, i.e., the ratio between the 
horizontal and the vertical dimensions of the 
vessel. 
For big aspect ratios ($a > 20$)
regular patterns appear at the onset of convection (typically 
hexagons in BM convection). For small $a$ ( $a < 10$) 
the arising patterns are determined by the geometry of 
the vessel. 
In a theoretical work  Rosenblat {\em et al.} studied the linear and 
weakly nonlinear problem with small aspect ratio for cylindrical 
\cite{Rosenblat1} and square \cite{Rosenblat2} containers. 
In the linear case buoyancy and thermocapillary forces were
considered. But the extension to the nonlinear regime was 
only made for pure thermocapillary effects.

There are important experimental contributions in 
these two regimes. For big aspect ratio the most interesting are 
due to Koschmieder \cite{kosbok}, Cerisier \cite{cer} and Schatz {\em et al.} 
\cite{swin}.
For small aspect ratio Koschmieder and Prahl \cite{koschi2} performed  
experiments in square and cylindrical containers near convective threshold.
Ondarcuhu {\em et al.} described a series of 
dynamical patterns that look very different in square \cite{ondar1} and in 
cylindrical containers \cite{ondar2}.

We study BM convection in a cylindrical container of 
small aspect ratio in the usual case when both, 
buoyancy and thermocapillarity forces, act together. 
In order to study the convective regime time dependent amplitudes 
of the unstable modes 
are considered. Bifurcation theory shows 
\cite{Golubitsky} that in a system with $O(2)$ 
symmetry but with the reflection symmetry with respect to the middplane 
broken (like in BM convection or in Rayleigh-B\'{e}nard 
convection under non-Boussinesq conditions) not 
only the modulus, but also the relative phases of the 
amplitudes must be taken into account. {\em Codimension two} (CT) 
points, where two modes become simultaneously unstable are specially 
interesting because they can lead to a resonant interaction with 
a characteristic 
dynamical behaviour. A similar case has been studied 
in convection in spherical cells (SO(3) symmetry) 
by Friedrich and Haken \cite{Fiddi}.

In Section 
2 we present the basic equations and the boundary 
conditions governing the dynamics of the 
system. In Section 3 we perform a linear analysis 
and compare our results with the experimental ones. 
Section 4 is devoted to the weakly nonlinear regime. 
In particular, we derive the amplitude 
equations for a 1:2 resonance, which were also derived 
for a model equation 
by means of bifurcation theory \cite{Dangelmayr} 
and studied in detail in ref. \cite{Proctor}. A physical 
interpretation of these 
results is made. Finally, we discuss the main conclusions 
in Section 5.
 
\section{Evolution equations and boundary conditions}

Consider a fluid layer of depth $d$ 
confined between a conducting plate heated from below 
and an upper surface open to the atmosphere, under 
a temperature difference $ \Delta T$. We 
assume that the fluid is Newtonian and that Boussinesq 
approximation holds. (We will follow the standard notation 
for the parameters of the fluid).

Then, in nondimensional 
variables (obtained dividing space, time, velocity 
and temperature by $d$, $d^2/\kappa$, $\kappa/d$ 
and $\Delta T$ respectively) the equations governing 
the dynamics of the fluid are:\par

\begin{equation}
\nabla\cdot{\bf v}=0.
\end{equation}\par
\begin{equation}
Pr^{-1}\left [\frac{\partial {\bf v}}{\partial t} + 
({\bf v}\cdot\nabla){\bf v} \right ]=-\nabla p + 
R T{\bf e_z} + \nabla^2 {\bf v}.\label{NS}
\end{equation}\par
\begin{equation}
\frac{\partial T}{\partial t} + {\bf v}\cdot\nabla T
=\nabla^2 T,\label{H.cond.}
\end{equation}\par
\noindent
where the Prandtl number $Pr=\nu/\kappa$ gives an estimation of the 
relative importance of thermal and mechanical 
dissipative processes. In this work we take the 
limit $Pr\rightarrow\infty$, equivalent to assume 
that velocity perturbations relax 
much faster than temperature perturbations. 
This limit allows to neglect both, the advective term 
and the temporal derivative of the velocity in 
the Navier-Stokes equation, so the velocity field 
follows {\em adiabatically} the variations of  
temperature. The Rayleigh number ($R=\alpha g 
d^3 \Delta T/\nu\kappa$) and the Marangoni number 
($M=\gamma d \Delta T/\rho \nu\kappa$) measure 
the relative importance of buoyancy and thermocapillarity 
forces. They are not independent but related by a 
constant $\Gamma=\gamma/\rho\alpha g d^2$ ($M=\Gamma R$). 
The limit $\Gamma\rightarrow 0$ corresponds to 
Rayleigh-B\'{e}nard convection while $\Gamma
\rightarrow\infty$ is reached when gravity forces 
are absent or the thickness of the fluid is very 
small ($R\sim d^2 M$). In this work we will analyse three
situations: a) $\Gamma=0.05$, corresponding to the experiments
in refs. \cite{ondar1}, \cite{ondar2}, b) $\Gamma=1$, for 
some situations analysed in ref. \cite{koschi2} and 
c) $\Gamma=100$, when surface tension is much 
greater that buoyancy effects, like in microgravity.
In order to determine completely the problem we 
must give also some boundary conditions (bc). The 
bottom is assumed to be rigid and conducting
\begin{equation}
{\bf v}=0,\;T=T_0\;\makebox[1cm]{at}\;z=0,
\end{equation}
while for the upper surface, assumed to be nondeformable 
(which is a good approximation for layers which are thick 
enough) and partially conducting we 
consider heat transfer and the so 
called Marangoni condition:
\begin{equation}
\frac{\partial u}{\partial z}+M\frac{\partial T}
{\partial r}=\frac{\partial v}{\partial z}+
\frac{M}{r}\frac{\partial T}{\partial \phi}=w=\frac{\partial T}
{\partial z}+Bi(T-T_{ref})=0\;\makebox[1cm]{at}\;z=1,
\end{equation}
where $T_{ref}$ is a reference temperature, 
$(u,v,w)$ are the components of the velocity 
in cylindrical coordinates $(r,\phi,z)$ and $Bi$ stands for 
the Biot number, a phenomenological parameter 
that accounts for the heat transfer between 
the fluid and the air. The limit $Bi\rightarrow 
\infty$ corresponds to a perfect conducting 
medium above the fluid while $Bi\rightarrow 0$ 
corresponds to an insulator material. We use 
for the rest of this paper
the value $Bi=0.1$ typical in experimental situations 
\cite{Mancini}.

In a finite container we must consider the 
horizontal extend by means of another nondimensional 
parameter, the aspect ratio $a$, defined as the ratio between 
the radius and the height of the cylindrical 
container ($a=R/d$). To complete the problem we need  
some bc's on the sidewall that depend on the geometry of the vessel. 
For example, adiabatic walls are described by the condition:
\begin{equation}
\frac{\partial T}{\partial r}=0\;\makebox[1cm]{at}\;r=a.
\end{equation}
In usual experiments \cite{koschi2,ondar2} plastic sidewalls, matching that 
bc are used. 
For the velocity, instead of the usual non-slip condition 
we take non-deformable walls on which the tangential vorticity 
vanishes (slippery walls)
\begin{equation}
u=\frac{\partial}{\partial r}(rv)=
\frac{\partial w}{\partial r}=0\;\makebox[1cm]{at}\;r=a,
\end{equation}
This bc is used because the problem become separable, so 
analytical methods can be used. Slippery "walls" are obtained in the separation among 
convective cells in a big aspect ratio system. 
Although this b.c. 
does not correspond to that used in experiments, the 
corresponding solutions are expected to reproduce qualitatively the 
experimental facts. 

Once the equations governing the dynamics
are given we can seek for solutions 
of these equations. In particular, we are 
interested in those solutions that appear 
when the trivial one (the conductive state) 
becomes unstable.

\section{Critical modes}

In this section we briefly recall the main results of the 
linear analysis of the equations and b.c.
The conductive state has velocity and 
temperature fields in the form:
\begin{equation}
{\bf v}_{cond}=0,\;T_{cond}=-z+T_0
\end{equation}\par
\noindent
and conserving only the linear terms we obtain:
\begin{eqnarray}
\nabla\cdot{\bf v}=0.\\
\frac{\partial {\bf v}}{\partial t}=Pr(-\nabla\pi+
R\theta{\bf e_z}+\nabla^2{\bf v}).\\
\frac{\partial \theta}{\partial t}=w+\nabla^2\theta,
\end{eqnarray}
with bc:
\begin{eqnarray}
{\bf v}=\theta=0\;&\makebox[1cm]{at}\;&z=0.\label{bcz0}\\
\frac{\partial u}{\partial z}+M\frac{\partial \theta}{\partial x}=
\frac{\partial v}{\partial z}+M\frac{\partial \theta}{\partial y}=
w=\frac{\partial \theta}{\partial z}+Bi\theta=0&\;\makebox[1cm]{at}&\;z=1.\\
\frac{\partial \theta}{\partial r}=u=\frac{\partial}{\partial r}(rv)=
\frac{\partial w}{\partial r}=0&\;\makebox[1cm]{at}\;&r=a.\label{bcra}
\end{eqnarray}\par
\noindent
Solutions of these equations can be written as: 
${\bf v}({\bf r},t)={\bf v}
({\bf r})e^{st}$, $\theta ({\bf r},t)=\theta ({\bf r})e^{st}$, 
$\pi ({\bf r},t)=\pi ({\bf r})e^{st}$, where $s$ 
is the growth rate. When marginal stability ($s=0$) 
holds, we have:
\begin{eqnarray}
\nabla\cdot {\bf v}=0.\label{Cont.2}\\
-\nabla\pi + R\theta{\bf e_z}+\nabla^2{\bf v}=0.\\
\nabla^2\theta+w=0.\label{H.cond.2}
\end{eqnarray}

Separation of variables allows to consider 
solutions of the system (\ref{Cont.2})-(\ref{H.cond.2}) 
with bc (\ref{bcz0})-(\ref{bcra}) given by:
\begin{equation}
\left .
\begin{array}{l}
u_{mij}(r,\phi,z)=(1/k_{mi})\cos(m\phi+\phi_m)J'_m(k_{mi}r)
DW_{mij}(z)\\
v_{mij}(r,\phi,z)=(-m/k^{2}_{mi}r)\sin(m\phi+\phi_m)J_m(k_{mi}r)
DW_{mij}(z)\\
w_{mij}(r,\phi,z)=\cos(m\phi+\phi_m)J_m(k_{mi}r)
W_{mij}(z)\\
\theta_{mij}(r,\phi,z)=\cos(m\phi+\phi_m)J_m(k_{mi}r)
\Theta_{mij}(z)
\end{array}
\right \},
\end{equation}
where $m=0,1,2,\dots$ is the azimuthal wavenumber, 
$J_m$ is the Bessel function of order $m$ and $i$  
is the radial wavenumber, which indexes the values 
$k_{mi}$ satisfying
the condition $J'_m(ka)=0$.
Due to symmetry there is no bc for the 
azimuthal equation, so the phase is not fixed and must be taken into 
account. This differs 
from the analysis of Rosenblat {\em et al} \cite{Rosenblat1}, 
in which the phase dynamics was not considered.   

The functions $W_{mij}(z)$, $\Theta_{mij}(z)$ are 
solutions of the system
\begin{eqnarray}
&&(D^2-k^2)^2 W-Rk^2\Theta=0.\label{Ecz}\\
&&(D^2-k^2)\Theta + W=0,
\end{eqnarray}
with bc
\begin{eqnarray}
&&W=DW=\Theta=0 \;\makebox[1cm]{at}\; z=0.\\
&&W=D\Theta+Bi\Theta=D^2W+Mk^2\Theta=0 \;\makebox[1cm]{at}\; z=1,
\label{bczz1}
\end{eqnarray}
where $D \equiv \frac{d}{dz}$. These equations have solutions for several values of 
$R$ denoted by $R_j(M, Bi, m, i,a)$, so the Rayleigh number
can be considered as the eigenvalue of the former 
system (and $j$ may be considered as a vertical wavenumber). 
The critical Rayleigh number ($R_c$) is 
the minimum eigenvalue for which equations 
(\ref{Ecz})-(\ref{bczz1}) have a solution. It 
corresponds to the onset of convection. Nevertheless 
we must notice that in a real system we do not 
fix $M$ but $\Gamma$. Therefore, in order to obtain the 
critical value $R_c$ we eliminate $M$ using the 
relation $M=\Gamma R$. Then, the critical Rayleigh 
number is:
\begin{equation}
R_c=\min_{mij} R_j(\Gamma,Bi,a,m,i)
\end{equation}\par
\noindent 
Once $R_c$ determined the rest of the 
eigenvalues $R_j(M_c,Bi,a,m,i)$ and the 
eigenfuntions $W_{mij}$, $\Theta_{mij}$ 
are calculated keeping the Marangoni 
number fixed $M_c=\Gamma R_c$. In Fig. 1 we plot
$R_c$ versus $a$ for the different modes for $\Gamma = 0.05$. 
For $\Gamma = 100$ we recover the results quoted in 
ref. \cite{Rosenblat1}. 
Experiments are in qualitative agreement
with these results, 
but a quantitative comparison is not possible 
mainly due to the idealized bc considered in our analysis.
It was noticed that modes 
with $m =3$ appear for $a = 5-6$ in experiments, 
a factor 2 bigger than our calculations \cite{ondar2}. 
Another discrepancy is that the mode 
$m = 0$ is observed in convective cells with 
$a < 1$ \cite{koschi2} a fact that cannot be 
explained theoretically. 
However, numerical simulations of 
the Rayleigh-Benard 
problem with realistic lateral bc 
show the same sequence as in the present paper 
\cite{Gershuni}. 

\section{Weakly nonlinear expansion}

We perform a Galerkin-Eckhaus expansion 
of the fields in terms of eigenfunctions 
of the linear problem with time-depending amplitudes. 
After inserting this expansion in the PDE's and 
projecting over the modes of the adjoint 
problem, an infinite-dimensional system of 
ODE's for those amplitudes is obtained. Then, 
a {\em center manifold reduction} is performed into  
the few 
modes that remain dynamically active 
after all transients have relaxed. 

Projecting the nonlinear system on the 
eigenfunctions of the adjoint system 
one arrives to (see Appendix) 
\begin{equation}
\left (1-\frac{R_c}{R}\right )\langle 
\theta^{*}_{mij}w\rangle +\frac{1}{\Gamma^2 R}\left 
(1-\frac{R^{*}_{mij}}{R_c}\right )\langle w^{*}_{mij}
\theta\rangle=\langle \theta^{*}_{mij}\frac
{\partial\theta}
{\partial t}\rangle+\langle \theta^{*}_{mij}
({\bf v}\cdot\nabla)\theta\rangle,\label{Ec.amp.} 
\end{equation}
where ${\bf v}^{*}_{mij}$ and $\theta^{*}_{mij}$ 
are the adjoint modes and $\langle \cdots \rangle$ 
denotes the average over the fluid volume:
\begin{equation}
\langle B\rangle\equiv \frac{1}{V}\int_{V} B\;
dV=\frac{1}{\pi a^2}\int^{1}_{0}dz\int^{a}_{0}r\;
dr\int^{2\pi}_{0}d\phi\;B
\end{equation}
The velocity and temperature 
fields are expanded in terms of the modes 
of the linear problem:
\begin{equation}
\left \{ \begin{array}{c} w\\ \theta \end{array}\right \}=
\sum_{mij}(A_{mij}(t)e^{im\phi}+\overline{A}_{mij}(t)
e^{-im\phi})J_m(k_{mi}r)\left \{ \begin{array}{c} W_{mij}(z)\\
 \Theta_{mij}(z)  \end{array} \right \}\label{amp.expan.}
\end{equation}
where $A_{mij}=R_{mij}e^{i\phi_m}$ is a complex 
amplitude with an indetermined phase $\phi_m$ included.
We restrict our calculations to the neigbourhood of a codimension 
two (CT) point where two modes arise simultaneously. 
In our case the spatially homogeneous amplitude equations 
for these two modes match the 
{\em normal form} of such a CT point with $O(2)$ 
symmetry  \cite{Golubitsky}:
\begin{eqnarray}
&&\dot{A}_l=p_lA_l+q_l\overline{A}_{l}^{m-1}A_{m}^{l}.\\
&&\dot{A}_m=p_mA_m+q_mA_{l}^{m}\overline{A}_{m}^{l-1},
\end{eqnarray}
where $p_j,q_j$ are functions of $|A_l|^2,|A_m|^2$ 
and $A_{l}^{m}\overline{A}_{m}^{l}+
\overline{A}_{l}^{m}A_{m}^{l}$ satisfying 
$p_j(0)=0,q_j(0)=0$. The last terms of 
these equations are called {\em resonant terms}. 
If $m+l-1\leq 3$ these resonant terms 
appear in the normal form. 

A {\em strong resonance} \cite{Dangelmayr} 
is obtained between the two modes 
(11) and (21) for $a = a_{CT}$ (see Fig. 1), where the 
value $a_{CT}$ slightly depends on $\Gamma$ 
( $ a_{CT} = 1.15 $ for  $\Gamma = 0.05$, and 
$ a_{CT} = 1.17$  for  $\Gamma = 1, 100$) and 
the corresponding critical Rayleigh number 
for this point is denoted as $R_{CT}$ ($R_{CT} = 
530$ for  $\Gamma = 0.05$, $R_{CT} = 80.5$ for  
$\Gamma = 1$, $R_{CT} = 0.894$ for  $\Gamma = 100)$. 
Under these conditions the normal form is:
\begin{eqnarray}
&&\dot{A}_1=\mu_1A_1+\alpha_1\overline{A}_1 A_2-
a_1A_1|A_1|^2-
b_1A_1|A_2|^2.\label{Ec.amp.1}\\
&&\dot{A}_2=\mu_2A_2-\alpha_2A_{1}^{2}-a_2A_2|A_2|^2-
b_2A_2|A_1|^2,\label{Ec.amp.2}
\end{eqnarray}
where we have taken $p_1=\mu_1-a_1|A_1|^2-b_1|A_2|^2+
{\cal O}(A^3)$, $p_2=\mu_2-a_2|A_2|^2-b_2|A_1|^2+{\cal O}(A^3)$, 
$q_1=\alpha_1+{\cal O}(A^2)$, $q_2=-\alpha_2+{\cal O}(A^2)$.

The form of Eqs. (\ref{Ec.amp.1})-(\ref{Ec.amp.2}) 
does not depend on the details of the system, 
but arises from the resonance and the symmetry properties.
The value of the coefficients depends on the specific 
problem and on the truncation order in the Galerkin-Eckhaus 
expansion. We consider the critical modes (11), (21) and a finite 
set of stable modes with lower growth rate, i.e.,  ((01), (31), (41), (12)), 
which are then eliminated using the center 
manifold reduction (see Appendix for details). The values 
of the coefficients under these approximations are given in 
Table~\ref{tab.fn.}. 

Eqs. (\ref{Ec.amp.1})-(\ref{Ec.amp.2}) have 
been studied by several authors \cite{Dangelmayr},
\cite{Proctor}. Here we summarize the 
main results. Using 
polar coordinates for the amplitudes $A_1= r_1
e^{i\phi_1}$ and $A_2= r_2 e^{i\phi_2}$ 
system (\ref{Ec.amp.1})-(\ref{Ec.amp.2}) 
is written 
\begin{eqnarray}
&&\dot{r_1}=\alpha_1 r_1 r_2 \cos \Phi+
\mu_1 r_1-a_1 r_1^3-b_1 r_1 r_2^2.\label{Ec.r_1}\\
&&\dot{r_2}=-\alpha_2 r_1^2\cos\Phi+
\mu_2 r_2-a_2 r_2^3-b_2 r_2 r_1^2.\\
&&\dot{\Phi}=(\alpha_2 \frac{r^2_1}{r_2} -
2\alpha_1 r_2)\sin\Phi,\label{Ec.Phi}
\end{eqnarray}
with $\Phi=2\phi_1-\phi_2$. (Rotational invariance 
allows to remove one of the phase equations).

There are three different types of steady-state 
solutions of Eqs. (\ref{Ec.r_1})-(\ref{Ec.Phi}):

1) {\em Pure modes} ($P_{\pm}$):
\begin{equation}
r_1=0,\; r_2^2=\frac{\mu_2}{a_{2}},\;
\Phi_+=0,\;\Phi_-=\pi.
\label{eq.pur.}
\end{equation}
\noindent
These solutions represent convection with only 
one mode and correspond to a temperature field in the  
form (see Eq. (\ref{amp.expan.})):
\begin{equation}
\theta(r,\phi)\propto\sqrt\frac{\mu_2}{a_2} 
\cos(2\phi)J_2(k_{21}r)\;\;\;\;(P_-)\\
\end{equation}
\noindent
For the sake of experimental comparison 
we plot the perturbation of the temperature  
field on the upper surface 
for the pure modes $(21)$ and $(11)$ in Fig. 2, although only the 
former one is solution of Eq. (\ref{eq.pur.}). 
The pure mode $(21)$ has two reflection ($Z_2$) symmetries and the
$(11)$ only the reflection respect to the diagonal that unites the 
upflow and downflow motions.
In experiments shadowgraphy and schlieren techniques 
are used to analyse the patterns \cite {ondar1}, \cite {ondar2}.  
In the corresponding images downflow motions appear as 
dark zones and upflow motions as bright zones. The mode 
$(21)$ must give images with a 
cold line (downflow) in a diagonal that joins two black zones, 
while the mode $(11)$ 
gives up and down
motions near the sidewalls on the two opposite sites of a diagonal.

Pure mode solutions exist provided $a_2\mu_2>0$ and 
lose their stability to {\em mixed modes} when
\begin{equation}
\mu_1>\frac{b_1\mu_2}{a_2}\pm \alpha_1\left 
(\frac{\mu_2}{a_2}\right )^{\frac{1}{2}},
\end{equation}
that we will denote line $\Pi$.

2) {\em Mixed modes} ($M_{\pm}$):
\begin{eqnarray}
&&0=\mu_1\pm \alpha_1 r_2-a_1{r_1}^2-b_1{r_2}^2,
\nonumber\\
&&0=\mu_2 r_2\mp \alpha_2 {r_1}^2-b_2{r_1}^2 r_2-a_2
{r_2}^3,\label{mixed}\\
&&\Phi_+=0,\;\Phi_-=\pi.\nonumber
\end{eqnarray}
\noindent
The two mixed modes represent mixed states 
of modes $(11)$ and $(21)$ (Eq. (\ref{amp.expan.})): 
\begin{equation}
\theta(r,\phi)\propto r_1\cos(\phi) J_1(k_{11}r)+
r_2\cos (2\phi) J_2(k_{21}r)\;\;(M_+)
\end{equation}
\noindent
Two examples are given in Fig. 3. 
Notice that these two modes mix in such a way that only one of the two $Z_2$ 
symmetries of the $(21)$ mode is broken, i.e, the angle between
these two modes is 0 ($M_+ $) or $\pi$ ($M_-$). 
The mixed mode 
$M_+$ undergoes a Hopf bifurcation on the 
line given by (\ref{mixed}) together with
\begin{equation}
a_1 {r_1}^2+a_2 {r_2}^2=\frac{\alpha_2 {r_1}^2}
{2 r_2},
\end{equation}
\noindent
which gives rise to standing waves (SW) 
characterized by $\dot{\Phi}=0$, but time-dependent 
amplitudes $r_1$ and $r_2$. We will 
denote $\Lambda$ the line in which this ocurs.

3) {\em Rotating waves} (RW): 
\begin{eqnarray}
&&{r_1}^2=2\frac{\alpha_1}{\alpha_2}{r_2}^2,\;
 {r_2}^2=\frac{2\mu_1+\mu_2}{\Delta},\\
&&\cos\Phi=\frac{1}{\alpha_1 r_2}
\frac{\mu_2[2(\alpha_1/\alpha_2)a_1+b_1]-
\mu_1[2(\alpha_1/\alpha_2)b_2+a_2]}{\Delta},
\label{cosPhi}
\end{eqnarray}
\noindent
with
\begin{equation}
\Delta\equiv 4\frac{\alpha_1}{\alpha_2}a_1+2b_1+
2\frac{\alpha_1}{\alpha_2}b_2+a_2.\label{Delta}
\end{equation}
\noindent
At first order in perturbations they can be represented as:
\begin{equation}
\theta(r,\phi)\propto\sqrt{\frac{2\mu_1+\mu_2}{\Delta}}
\left (\sqrt{2\frac{\alpha_1}{\alpha_2}}\
cos(\phi+\dot{\phi_1}t)J_1(k_{11}r)+\cos(2(\phi+
\dot{\phi_1}t)-\Phi)J_2(k_{21}r)\right )
\;(RW)
\end{equation}
\noindent
with $\Delta$, $\dot{\phi_1}$ and $\Phi$ given by 
Eqs. (\ref{Delta}), (\ref{Ec.ph.vel.}) and (\ref{cosPhi}). 

In this state the phases of the two modes $\phi_1$ and $\phi_2$ 
are variable but the 
relative phase $\Phi$ is constant and different
from zero. So, the remaining $Z_2$ symmetry of the $(11)$
mode is broken, as it is shown in the two examples given in Fig. 4.
These rotating waves exist provided that $(2\mu_1+
\mu_2)\Delta>0$ and $|\cos \Phi|\leq 1$, i.e.,
\begin{equation}
\left [\left (2\frac{\alpha_1}{\alpha_2}a_1+
b_1\right )\mu_2-\left (2\frac{\alpha_1}
{\alpha_2}b_2+a_2\right )\mu_1\right ]^2
\leq\alpha_{1}^{2}(2\mu_1+\mu_2)\Delta,
\end{equation}
\noindent
and bifurcate off $M_{\pm}$ at $|\cos \Phi|=1$ (line $\Omega$).

The pattern of these RW rotates with a phase velocity
\begin{equation}
\dot{\phi_2}=2\dot{\phi_1}=-2\alpha_1 r_2\sin\Phi=
-2\alpha_1\sqrt{\frac{2\mu_1+\mu_2}{\Delta}\left 
(1-\frac{(\beta\mu_2-\gamma\mu_1)^2}{\alpha_{1}^{2}
\Delta(2\mu_1+\mu_2)}\right )},\label{Ec.ph.vel.}
\end{equation}
\noindent
where $\beta\equiv 2(\alpha_1/\alpha_2)a_1+b_1$ and $\gamma\equiv 
a_2+2(\alpha_1/\alpha_2)b_2$. In Fig.~5 the phase velocity 
(\ref{Ec.ph.vel.}) is plotted as a function of 
$R$ for several values of the aspect ratio $a$. 
This has been calculated from the equation 
\begin{equation}
\mu_m=\frac{1}{\tau_m}\langle \theta^{*}_{m11}w_{m11}\rangle \left 
(1-\frac{R_{c_m}(a)}{R} \right ),\;m=1,2,\label{Ec.muep.}
\end{equation}
\noindent
that relates the control parameters $\mu_1$, 
$\mu_2$ with the physically relevant 
nondimensional numbers $R$ and $a$. 

The RW can become unstable to {\em modulated waves} (MW) 
with time dependent amplitudes and phase velocities 
$\dot{\phi_1},\dot{\phi_2}\not=0$. 

\subsection{Heteroclinic orbit}

When condition
\begin{equation}
0>\mu_1-\frac{b_1\mu_2}{a_2}>-\alpha_1\left (\frac
{\mu_2}{a_2}\right )^{\frac{1}{2}}\label{Eq.heter.}
\end{equation}
\noindent
is fulfilled, a stable {\em heteroclinic orbit} exists. 
We gathered in Fig. 6a the form of the patterns that 
may appear in four representative points of that orbit. 
Usually a point of this orbit corresponds to a mixture 
of the modes (21) and (11), but when it reaches the $r_1$-axis 
the phase suddenly 
jumps in $\pi$. Therefore, the amplitude of the mode (21) 
reverses sign. When $r_1 \simeq 0$ there is a pure mode 
(21) that varies its phase in $\pi/2$. Experimentally 
this heteroclinic orbit would lead to patterns evolving 
from one pure mode to another with a phase changed in
$\pi/2$.
This orbit is structurally stable to changes in the parameters 
$\alpha_i, a_i, b_i$. But it is interesting to notice that slight 
changes in the form of Eqs. (\ref{Ec.amp.1})-(\ref{Ec.amp.2}) 
(imperfect symmetry) destroy the heteroclinic orbit that 
degenerates into a limit cycle. In fact numerical 
noise suffices to convert the 
heteroclinic orbit into a periodic orbit in numerical simulations 
of Eqs.(\ref{Ec.amp.1})-(\ref{Ec.amp.2}) shown in Fig. 6b. Small 
thermal noise, cell imperfections, etc, unavoidable in real 
experiments will prevent the formation of the heteroclinic 
orbit which is replaced by periodic motions between modes
(11) and (21). 

In Fig. 7 the bifurcation diagram is plotted as 
function of $\mu_1$, $\mu_2$ for $\Gamma = 0.05$. 
This diagram can be plotted in the plane ($R$, $a$) as shown 
in Fig. 8. In Fig. 9 the stability diagrams for $\Gamma = 1$ 
and $\Gamma = 100$ are represented. The quotations in these 
figures correspond to the 
stability regions of the different solutions. From Figs. 7 and 8 
it can be deduced that for $a > a_{CT}$ the dynamics begins
with a mixed mode $(M_-)$ that changes into a rotating wave 
$(RW)$ when the  
heating is increased. For $a < a_{CT}$ convection starts
in a pure mode $(P)$ that bifurcates to a mixed mode $(M_+)$.
This is replaced by a heteroclinic orbit $(H)$ when the 
control parameter is rised. Notice that $(H)$
is stable in a full region of parameter space.
Therefore we hope that a periodic dynamics that arises from this 
heteroclinic connection
would be reachable in experiments with suitable  
conditions.

\section{Conclusions}

We performed an analysis of BM 
convection in a cylindrical
container with small aspect ratio. For the linear
 problem we calculated the 
marginal curve $R_c(a)$ and the unstable modes 
as a function of the aspect ratio $a$ for several values 
of the ratio between surface tension and buoyancy 
effects ($\Gamma$). These results have been 
obtained for idealized 
lateral boundary conditions (slippery walls). Therefore, they 
allow only for qualitative comparison with 
experiments. The sequence of modes and the 
critical Rayleigh number compares quite well 
with experimental findings. However, the critical 
aspect ratios calculated are smaller than those 
observed in experiments. This has also been obtained in the 
case of BM convection in a square container \cite{Dauby}.  
A possible explanation is 
that the viscous boundary layer in real non-slippery walls
rises the aspect ratio respect to an idealized slippery one.

The nonlinear analysis is performed near a 
codimension two (CT) point 
where two stationary pure modes are simultaneously unstable. We choose 
a situation where a resonance 2:1 between 
modes (11) and (21) appears. A 
center manifold reduction is done to determine 
the coefficients of the normal form in 
the neighbourhood of that point.
Depending on the parameter values the 
stationary solutions are: 
a) a pure mode (P) (with two reflection symmetries),
mixed modes (M) (with only one $Z_2$ 
symmetry), rotating waves (RW)
(without reflection symmetry)
and a heteroclinic orbit (H). We calculate 
the dependence of rotation 
velocity of RW as a function of the order parameter. 
The heteroclinic orbit is quite robust 
and appears as a stable 
solution in a big region of the 
parameter space (see Figs. 8, 9).  
In an experiment this orbit will 
degenerate into a periodic alternancy between two 
modes that could be detected in a wide region in the 
parameter space. We hope that new experiments will 
confirm the presence of a heteroclinic conection 
and of rotating waves in this system.\\[30pt] 

\leftline{\Large{{\bf Acknowledgements}}}

\vskip25pt

We acknowledge fruitful comments and discussions with P. Dauby 
and G. Lebon (Liege, Belgium), G. Mindlin (Buenos Aires, Argentina), 
H. Mancini, A. Garcimart\'{\i}n and D. Maza (Pamplona). This work has been
partilly funded by PIUNA (Universidad de Navarra) and by the
DGICYT (Spanish Government) under grant PB94-0708, and with an
European Union contract ERBCHRXT940546. B.E. 
acknowledges the Basque Government for a fellowship (BFI95.035) and D.K.  
the kind hospitality of the Universidad de Navarra.

\newpage
\renewcommand{\theequation}{A.\arabic{equation}}
\renewcommand{\thesection}{Appendix}
\setcounter{equation}{0} 
\setcounter{section}{0}
\section{}

The adjoint linear problem is given by the 
equations
\begin{eqnarray}
&&\nabla^2{\bf v}^*-\nabla\pi^*+\theta^*{\bf e}_z=
{\bf 0}.\\
&&\nabla\cdot{\bf v}^*=0.\\
&&\nabla^2\theta^*+R^* w^*=0,
\end{eqnarray}
with bc
\begin{eqnarray}
&&u^*=v^*=w^*=\theta^*=0,\;\makebox[1cm]{at}\;z=0,\\
&&w^*=\frac{\partial u^*}{\partial z}=
\frac{\partial v^*}{\partial z}=
\frac{\partial \theta^*}{\partial z}+
Bi\theta^*+M\frac{\partial w^*}{\partial z}=0,\;
\makebox[1cm]{at}\;z=1,\\
&&u^*=\frac{\partial}{\partial r}(rv^*)=
\frac{\partial w^*}{\partial r}=0,\;\makebox[1cm]{at}\;r=a.
\end{eqnarray}

This problem is separable and the eigenvalues 
$R^{*}_{mij}$ are the same of those of the 
system (\ref{Cont.2})-(\ref{H.cond.2}). The 
eigenfunctions are:
\begin{equation}
\left .
\begin{array}{l}
u^{*}_{mij}=k_{mi}\cos(m\phi)J'_{m}(k_{mi}r)
DW^{*}_{mij}(z).\\
v^{*}_{mij}=-(m/r)\sin(m\phi)J_m(k_{mi}r)
DW^{*}_{mij}(z).\\
w^{*}_{mij}=k^{2}_{mij}\cos(m\phi)J_m(k_{mi}r)
W^{*}_{mij}(z).\\
\theta^{*}_{mij}=\cos(m\phi)J_m(k_{mi}r)
\Theta^{*}_{mij}(z),
\end{array}
\right \}
\end{equation}
with $W^{*}_{mij}$ and $\Theta^{*}_{mij}$ 
solutions of the system
\begin{eqnarray}
&&(D^2-k^2)\Theta^*+Rk^2W^*=0.\\
&&(D^2-k^2)^2W^*-\Theta^*=0,
\end{eqnarray}
with bc
\begin{eqnarray}
&&\Theta^*=W^*=DW^*=0, \;\makebox[1cm]{at} \;z=0.\\
&&W^*=D^2W^*=D\Theta^*+Bi\Theta^*+k^2M_cDW^*=0,
\; \makebox[1cm]{at} \;z=1.
\end{eqnarray}

These eigenfunctions are introduced to obtain 
\begin{eqnarray}
\lefteqn{\left (1-\frac{R_c}{R}\right )
\langle \theta^{*}_{mij}w \rangle + \frac{1}
{\Gamma^2 R}\left (1-\frac{R^{*}_{mij}}{R_c}\right )
\langle w^{*}_{mij}\theta \rangle  }\nonumber\\
&&=\langle \theta^{*}_{mij}\frac{\partial 
\theta}{\partial t}+\frac{1}{Pr}{\bf v}^{*}_{mij}
\cdot\frac{\partial {\bf v}}{\partial t}\rangle + 
\langle \theta^{*}_{mij}({\bf v}\cdot\nabla)
\theta\rangle +
\frac{1}{Pr}\langle {\bf v}^{*}_{mij}\cdot
({\bf v}\cdot\nabla){\bf v}\rangle 
\end{eqnarray}
\noindent
that in the limit $Pr\rightarrow 
\infty$ leads to  
\begin{equation}
\left (1-\frac{R_c}{R}\right )\langle 
\theta^{*}_{mij}w\rangle +\frac{1}{\Gamma^2 R}\left 
(1-\frac{R^{*}_{mij}}{R_c}\right )\langle w^{*}_{mij}
\theta\rangle=\langle \theta^{*}_{mij}\frac
{\partial\theta}
{\partial t}\rangle+\langle \theta^{*}_{mij}
({\bf v}\cdot\nabla)\theta\rangle.
\end{equation}

Introducing expression (\ref{amp.expan.}) 
for the temperature and velocity fields, 
we obtain an infinite set of ODE's for the 
amplitudes. In order to perform 
a center manifold 
reduction we must take into account the critical modes and the set of 
stable modes that appear as cuadratical interaction of these ones (so they 
give thirth order terms in the center manifold equations). This is, the set: 
{$(0i), (1j), (2j), (3i), 
(4i)$},$i=1,2,\dots, j=2,3,\dots$. Since this is again 
an infinite set we will retain only 
those modes with the smallest eigenvalues $R^{*}_{mij}$ 
(see Table 2). Moreover we 
have checked that only modes $(01), (31), (41), (12)$ give 
an estimable contribution to 
the coefficients of the normal form. With this, the following amplitude equations are obtained:
\begin{equation}
\left.
\begin{array}{l}
\tau_{01}\dot{A}_{01}=\tilde{\epsilon}_{01} A_{01}+
\alpha_{010101}(A_{01}^{2}+|A_{01}|^2)+
\frac{\alpha_{011111}}{2}|A_{11}|^2+
\frac{\alpha_{011212}}{2}|A_{12}|^2+
\frac{\alpha_{012121}}{2}|A_{21}|^2\\
\makebox[1.5cm]{}+\frac{\alpha_{013131}}{2}|A_{31}|^2+
\frac{\alpha_{014141}}{2}|A_{41}|^2.\\

\tau_{11}\dot{A}_{11}=\tilde{\epsilon}_{11}A_{11}+
\alpha_{111121}\overline{A}_{11}A_{21}+
\alpha_{111221}\overline{A}_{12}A_{21}+
\alpha_{111101}(A_{11}A_{01}+
A_{11}\overline{A}_{01})\\
\makebox[1.5cm]{}+\alpha_{111201}(A_{12}A_{01}+
A_{12}\overline{A}_{01})+\alpha_{112131}\overline{A}_{21}A_{31}+
\alpha_{113141}\overline{A}_{31}A_{41}.\\

\tau_{21}\dot{A}_{21}=\tilde{\epsilon}_{21}A_{21}+
\alpha_{211111}A_{11}^{2}+\alpha_{211212}A_{12}^{2}+
\alpha_{211131}\overline{A}_{11}A_{31}+
\alpha_{211231}\overline{A}_{12}A_{31}\\
\makebox[1.5cm]{}+\alpha_{212101}(A_{21}A_{01}+
A_{21}\overline{A}_{01})+\alpha_{212141}\overline{A}_{21}A_{41}.\\

\tau_{31}\dot{A}_{31}=\tilde{\epsilon}_{31}A_{31}+
\alpha_{311121}A_{11}A_{21}+\alpha_{311221}A_{12}A_{21}+
\alpha_{311141}\overline{A}_{11}A_{41}+
\alpha_{311241}\overline{A}_{12}A_{41}\\
\makebox[1.5cm]{}+\alpha_{313101}(A_{31}A_{01}+A_{31}\overline{A}_{01}).\\

\tau_{41}\dot{A}_{41}=\tilde{\epsilon}_{41}A_{41}+
\alpha_{414101}(A_{41}A_{01}+A_{41}\overline{A}_{01})+\alpha_
{411131}A_{11}A_{31}+\alpha_{411231}A_{12}A_{31}\\
\makebox[1.5cm]{}+\alpha_{412121}A_{21}^{2}.\\

\tau_{12}\dot{A}_{12}=\tilde{\epsilon}_{12}A_{12}+
\alpha_{121121}\overline{A}_{11}A_{21}+
\alpha_{121221}\overline{A}_{12}A_{21}+
\alpha_{121101}(A_{11}A_{01}+
A_{11}\overline{A}_{01})\\
\makebox[1.5cm]{}+\alpha_{121201}(A_{12}A_{01}+
A_{12}\overline{A}_{01})+\alpha_{122131}\overline{A}_{21}A_{31}+
\alpha_{123141}\overline{A}_{31}A_{41}.\\
\end{array}
\right \}
\label{ecs.amp.2}
\end{equation}
with
\begin{eqnarray}
&&\tau_{mn}=\langle\theta^{*}_{mn1}
\theta_{mn1}\rangle,\;m=1\dots 4,\;n=1,2,\\
&&\tilde{\epsilon}_{i1}=\langle\theta^{*}_{i11}w_{i11}
\rangle\left (1-\frac{R_{c_i}}{R}\right )\equiv
\langle\theta^{*}_{i11}w_{i11}\rangle 
\epsilon_{i1},\;i=1,2\;\;\;(\epsilon_i \equiv 1-\frac{R_{c_i}}{R} ),\\
&&\tilde{\epsilon}_{jk}=\langle\theta^{*}_{jk1}
w_{jk1}\rangle\left (1-\frac{R_c}{R}\right )+
\frac{\langle w^{*}_{jk1}\theta_{jk1}\rangle}
{\Gamma^2 R}\left (1-\frac{R^{*}_{jk1}}
{R_c}\right )\simeq\frac{\langle w^{*}_{j11}\theta_{j11}
\rangle}{\Gamma^2 R_c}\left (1-\frac{R^{*}_{j11}}
{R_c}\right ),\\
&&\makebox[1.5cm]{}\mbox{if}\; R\simeq R_c\;(j=0,3,4,\;k=1\;\mbox{or}\;j=1,k=2),\\
&&\alpha_{ijklmn}=\langle\theta^{*}_{ij1}
({\bf v}_{kl1}\cdot\nabla)\theta_{mn1}\rangle
\end{eqnarray}
After eliminating the modes (01), (31), (41), (12) 
with a center manifold reduction, we obtain 
the equations for the critical modes (11), (21):

\begin{equation}
\begin{array}{l}
\tau_{11}\dot{A}_{11}=\tilde{\epsilon}_{11}A_{11}+\alpha_{111121}A_{21}\overline
{A}_{11}-\frac{\alpha_{111101}
\alpha_{011111}}{\tilde{\epsilon}_{01}}A_{11}|A_{11}|^2-(\frac
{\alpha_{110111}\alpha_{012121}}{\tilde{\epsilon}_{01}}\\
\makebox[1.5cm]{}+\frac{\alpha_{112131}\alpha_{311121}}{\tilde{\epsilon}_{31}}+\frac
{\alpha_{111221}\alpha_{121121}}{\tilde{\epsilon}_{12}})
A_{11}|A_{21}|^2.\\
\tau_{21}\dot{A}_{21}=\tilde{\epsilon}_{21}A_{21}+\alpha_{211111}
A_{11}^{2}-
(\frac{\alpha_{212101}\alpha_{012121}}{\tilde{\epsilon}_{01}}+
\frac{\alpha_{212141}\alpha_{412121}}{\tilde{\epsilon}_{41}})
A_{21}|A_{21}|^2\\
\makebox[1.5cm]{}-(\frac{\alpha_{211131}\alpha_{311121}}{\tilde{\epsilon}_{31}}+
\frac{\alpha_{212101}\alpha_{011111}}{\tilde{\epsilon}_{01}}+
\frac{\alpha_{211112}\alpha_{121121}}{\tilde{\epsilon}_{12}})
A_{21}|A_{11}|^2.\nonumber
\end{array}
\end{equation}
In Table~\ref{tab.rvc} the values of these coefficients are given for $\Gamma=0.05$. 
Dividing these by $\tau_{11}$ and
$\tau_{21}$ respectively, we obtain the coefficients given in 
Table~\ref{tab.fn.}.

\newpage
\begin{center}
{\Large{\bf Tables}}
\end{center}

\vspace{2cm} 

\begin{center}
\begin{table}[h]
\begin{displaymath}
\begin{array}{|c|c||c|c||c|c|} \hline
\multicolumn{2}{|c||}{\Gamma=0.05} & \multicolumn{2}{c||}{\Gamma=1} & 
\multicolumn{2}{c|}{\Gamma=100}\\ \hline
\mu_1=.1947\epsilon_1 &  \mu_2=.3599\epsilon_2 & 
\mu_1=0.06654\epsilon_1 &  \mu_2=0.09771\epsilon_2 &
\mu_1=0.05381\epsilon_1 &  \mu_2=0.09708\epsilon_2 \\ \hline
\alpha_1=.8376 & \alpha_2=.4069 &
\alpha_1=0.1842 & \alpha_2=0.04952 &
\alpha_1=0.03681 & \alpha_2=0.002350\\ \hline
a_1=17.45 & a_2=6.637 & 
a_1=2.130 & a_2=1.141 &
a_1=0.006209& a_2=0.003554\\ \hline
b_1=10.16 & b_2=28.36 &
b_1=1.603 & b_2=4.655 &
b_1=0.004855 & b_2=0.01384\\ \hline
\end{array}
\end{displaymath}												
\caption{Coefficients of the normal form for several values of $\Gamma$.
\label{tab.fn.}}
\end{table}
\end{center} 

\vspace{2cm}

\begin{center}
\begin{table}[h]
\begin{displaymath}
\begin{array}{|c|c|c|c|}        \hline
R^{*}_{111}=530.3 & R^{*}_{211}=530.3 & R^{*}_{011}=696.4 & R^{*}_{311}=812.2 \\ \hline
R^{*}_{411}=1326  & R^{*}_{121}=1332  & R^{*}_{221}=2403  & R^{*}_{321}=4039  \\ \hline
\end{array}
\end{displaymath}
\caption{Eigenvalues of the adjoint problem.
\label{Eigen.}}
\end{table}
\end{center} 

\vspace{2cm}

\begin{center}  
\begin{table}[h]
\begin{displaymath}
\begin{array}{|c|c|c|c|c|}        \hline
\tilde{\epsilon}_{01}=-3.102 &\tilde{\epsilon}_{11}=1.592\epsilon_{11} &
\tilde{\epsilon}_{21}=1.473\epsilon_{21} &
\tilde{\epsilon}_{31}=-.4902 &\tilde{\epsilon}_{41}=-1.168\\ \hline  
\tilde{\epsilon}_{12}=-3.889 &
\tau_{11}=8.179 & 
\tau_{21}=4.093 &\alpha_{012121}=7.750 & 
\alpha_{011111}=34.68\\ \hline 
\alpha_{111101}=-12.77 & \alpha_{112131}=9.211 &
\alpha_{112111}=6.851 &
\alpha_{211111}=-1.666 &\alpha_{211131}=4.255 \\ \hline
\alpha_{212101}=-8.267 &
\alpha_{311121}=-2.726 &\alpha_{412121}=-1.081 &
\alpha_{212141}=7.040 & \alpha_{211112}=-7.594 \\ \hline
\alpha_{111121}=6.851 & \alpha_{111221}=-11.545 &
\alpha_{121121}=5.685 & & \\ \hline
\end{array}
\end{displaymath}
\caption{Coefficients used in center 
manifold reduction for $\Gamma=0.05$.\label{tab.rvc}}
\end{table}
\end{center}  

\newpage

\begin{center}
{\Large{\bf Figure Captions.}}\par
\end{center}
\vspace{1cm} 

{\bf Figure 1}: Marginal curve representing the 
critical Rayleigh number $R_c$ as a function of 
the aspect ratio $a$. The indexes $(mi)$ stand for the 
azimuthal (m) and radial (i) wavenumbers. Only the modes with 
vertical wavenumber $j=1$ are represented 
because we have assumed the ordering $R_{mi1}<R_{mi2} \dots$ \par

{\bf Figure 2}: Pure modes a) $(11)$ and 
b) $(21)$. Only the second one appears as a solution 
of Eq. \ref{eq.pur.}. (Point a in Fig. 8). \par

{\bf Figure 3}: Mixed mode a) $M_+$, b) $M_-$. (Points b and c in Fig. 8).\par

{\bf Figure 4}: Rotating waves for a) $\Phi=2.0$ and b) $\Phi=2.25$. (Points 
d and e in Fig. 8).\par

{\bf Figure 5}: Phase velocity $\dot{\phi_2}$ (in absolute value) as a 
function of 
Rayleigh number $R$ when a) $a=1.00$, b) $a=1.15$, 
c) $a=1.30$. The solid line corresponds to the case when the 
$RW$ are stable and the doted line when they are unstable. 
These $RW$ appear as a supercritical bifurcation. From Eqs. 
(\ref{Ec.ph.vel.}) and (\ref{Ec.muep.}) it can be seen that, 
for $\epsilon\ll 1$, $\dot{\phi_2}\propto \epsilon^{\frac{1}{2}}$.
($\epsilon\equiv \frac{R-R_c}{R_c}$).\par

{\bf Figure 6}: a) Heteroclinic orbit. 
b) Temporal series of the amplitude $r_1$ when 
the system is in the former orbit. After a transient, the 
heteroclinic conexion merges into a 
limit cycle due to the numerical noise.\par

{\bf Figure 7}: Bifurcation diagram in the plane 
$\mu_1$, $\mu_2$. Parentheses indicate unstable solutions. 
The lines $\Sigma_1$ and $\Sigma_2$ represent the 
limits of existence of the heteroclinic orbit, as given by Eq. 
(\ref{Eq.heter.}). (For further explanations see the text).
       
{\bf Figure 8}: Stability diagrams in the plane ($R$, $a$). Only the
stable solutions are represented. When more than one solution is stable, 
that with the smallest basin of attraction is represented in parenthesis. 
The points a, b, c, d, e, f correspond to the solutions shown in Figs. 2b, 3a, 
3b, 4a, 4b and 6, respectively.\par

{\bf Figure 9}: Stability diagram in $(R, a)$ space, for 
a) $\Gamma = 1$ and b) $\Gamma = 100$. The stability region for the 
heteroclinic orbit (H) increses as $\Gamma$ increases.

\newpage

\epsfig{file=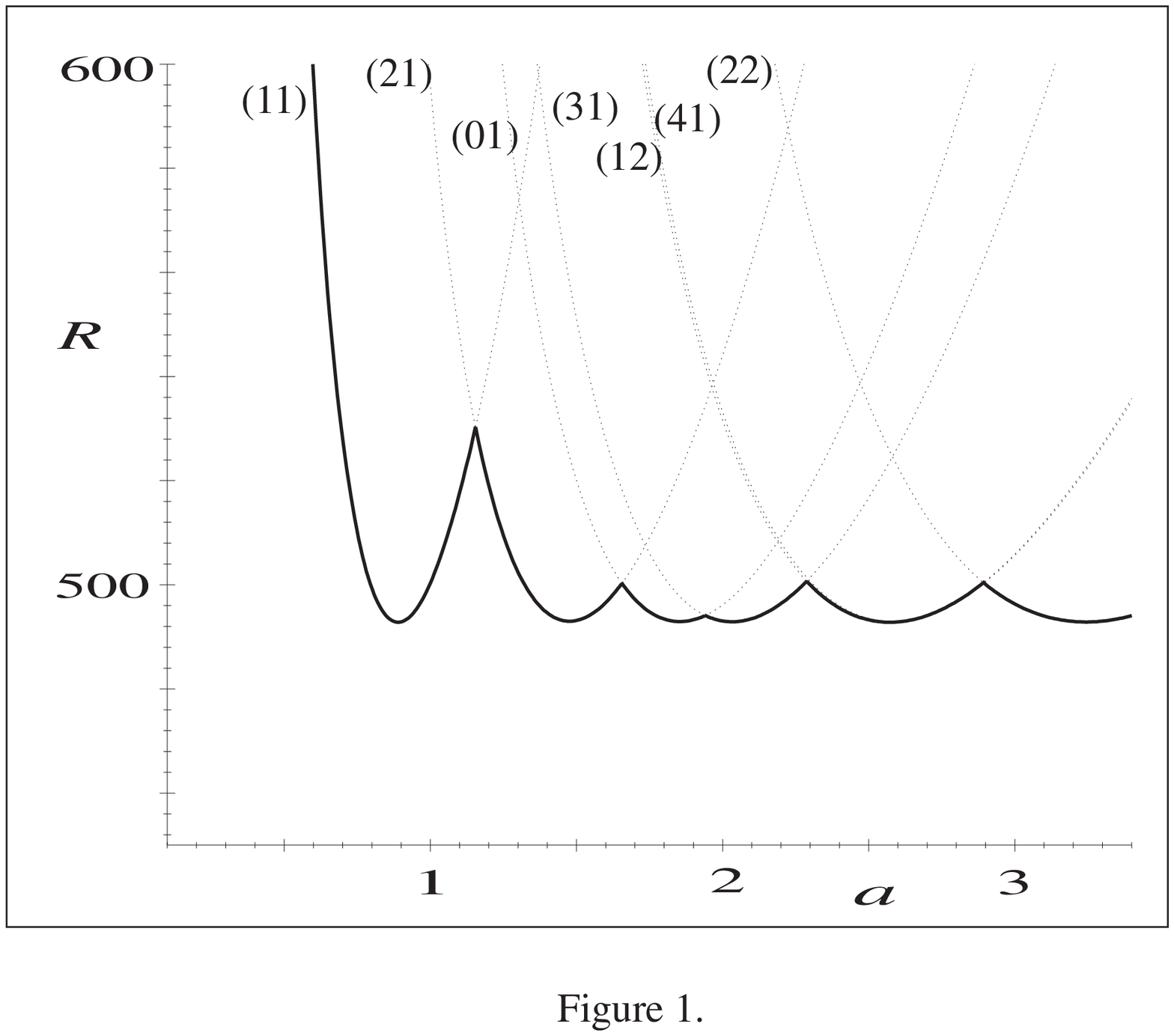}

\newpage

\hspace{2cm}\epsfig{file=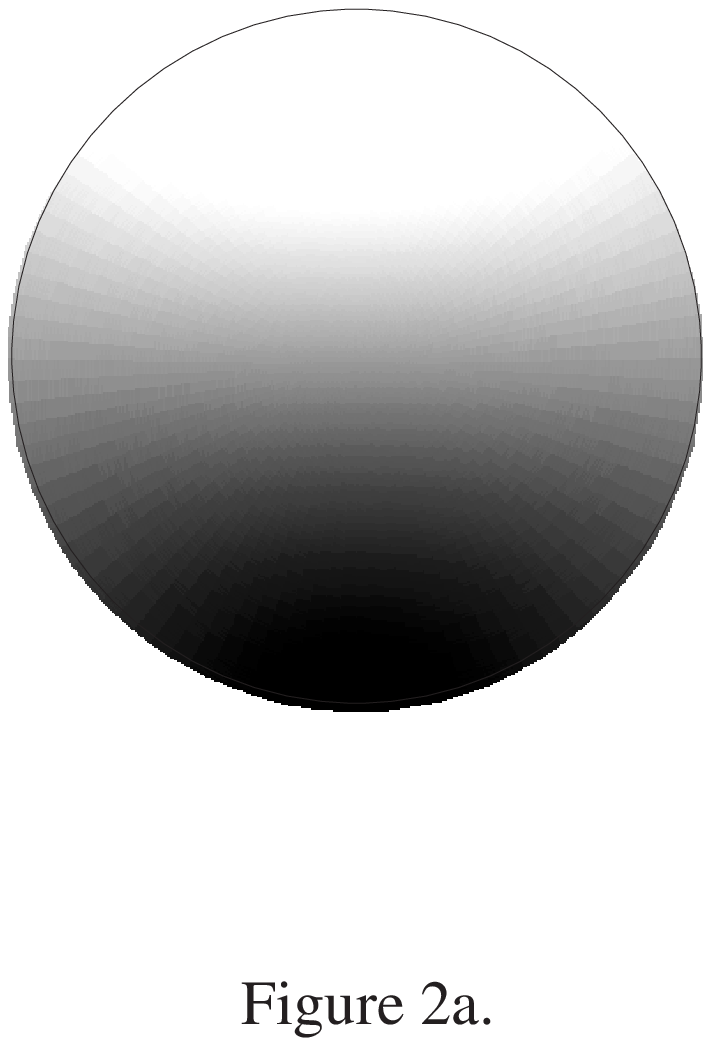}

\newpage

\hspace{2cm}\epsfig{file=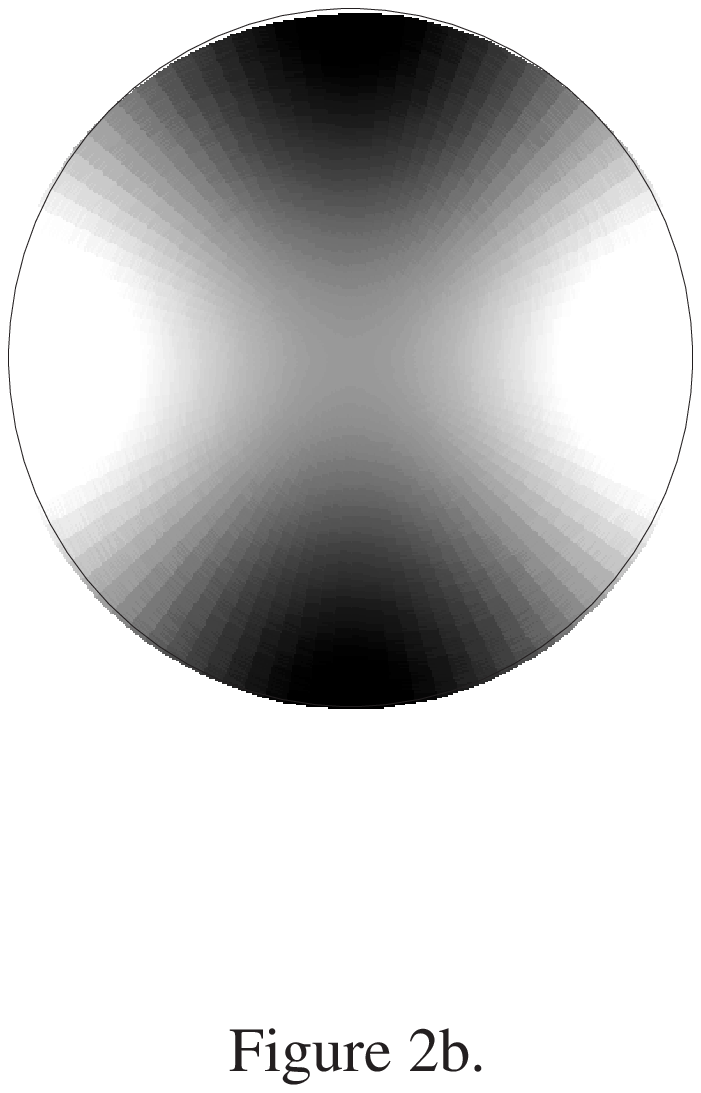}

\newpage

\hspace{2cm}\epsfig{file=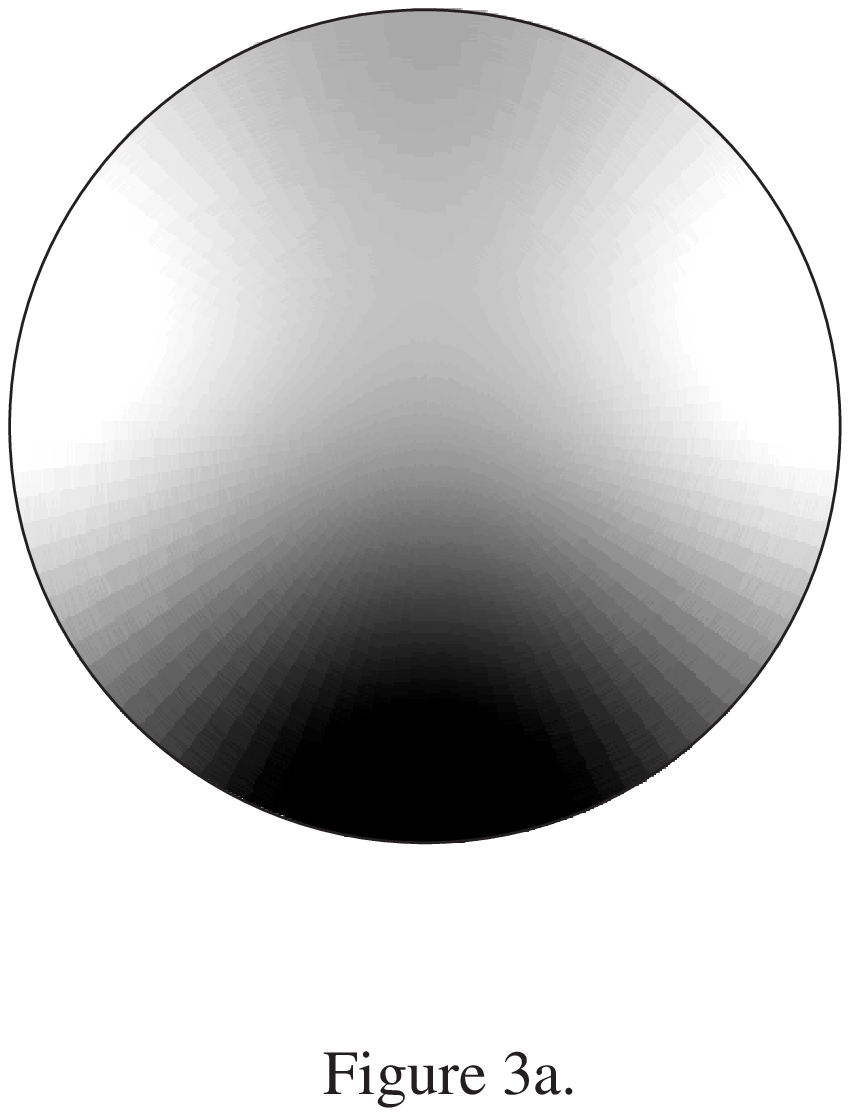}

\newpage

\hspace{2cm}\epsfig{file=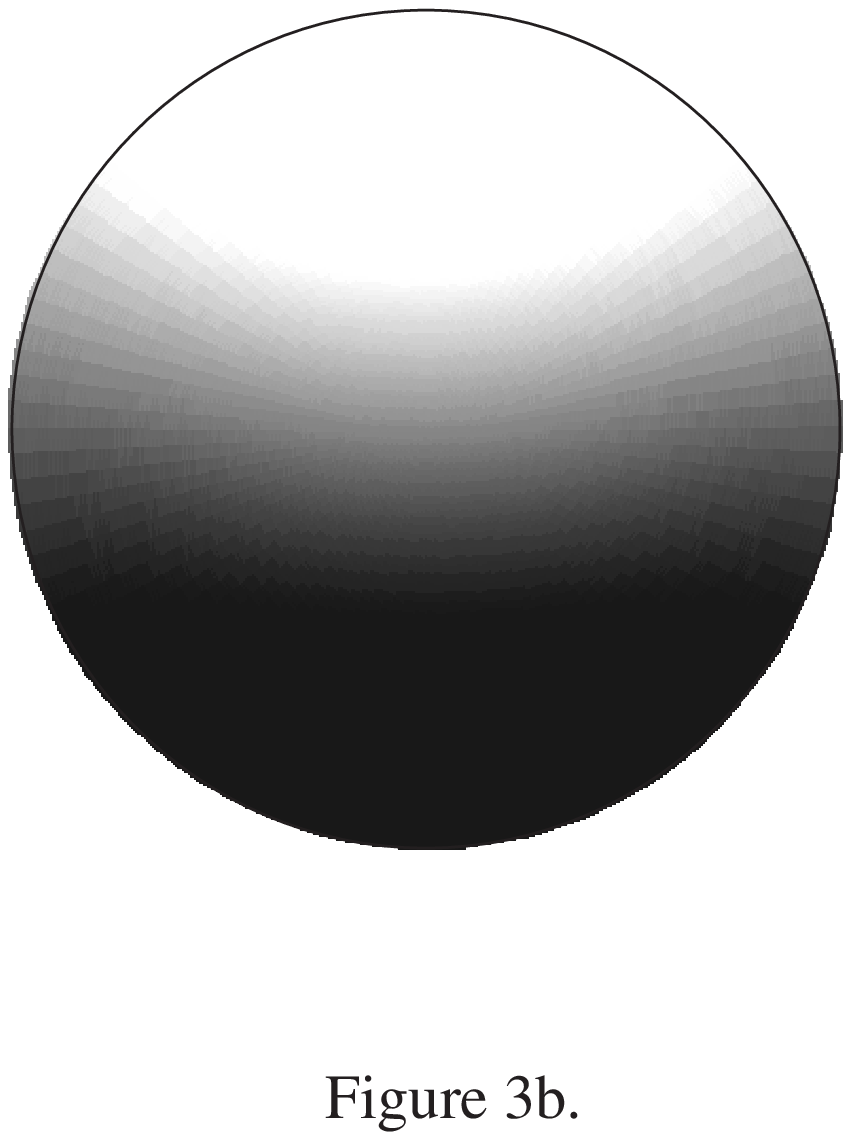}

\newpage

\hspace{2cm}\epsfig{file=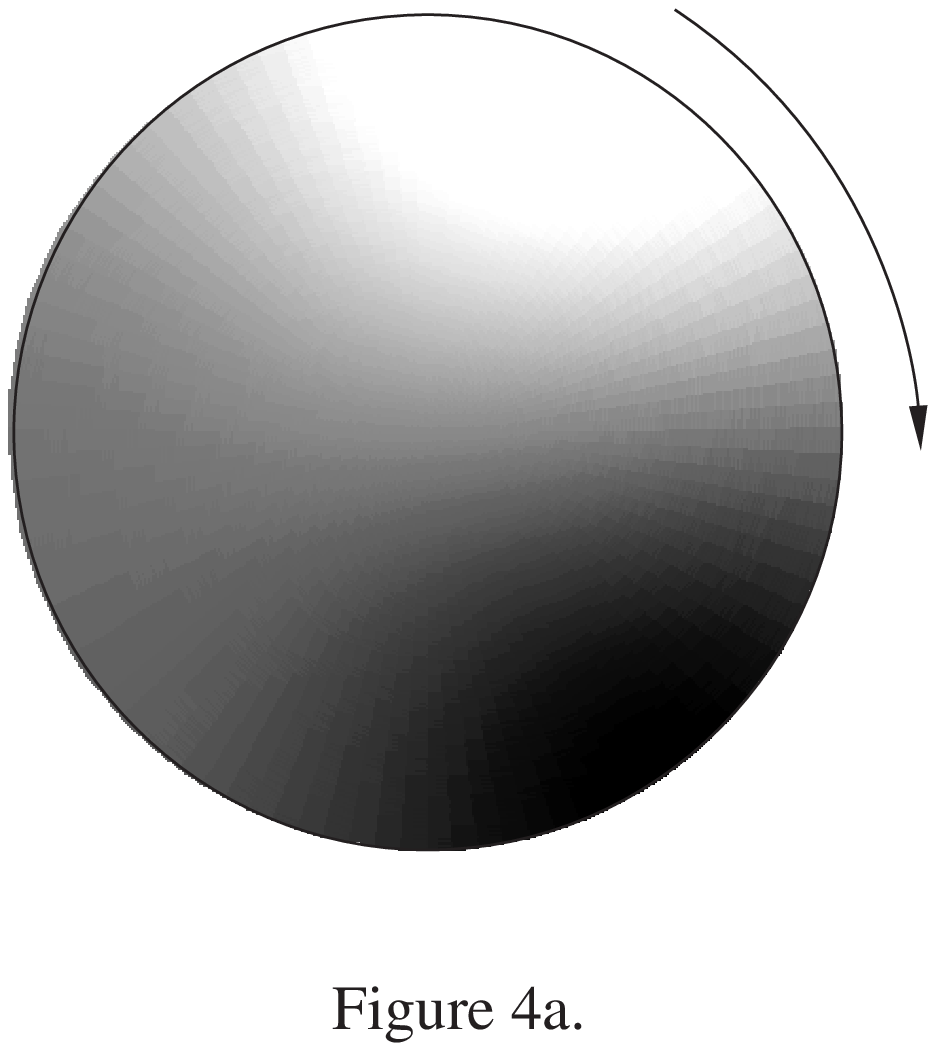}

\newpage

\hspace{2cm}\epsfig{file=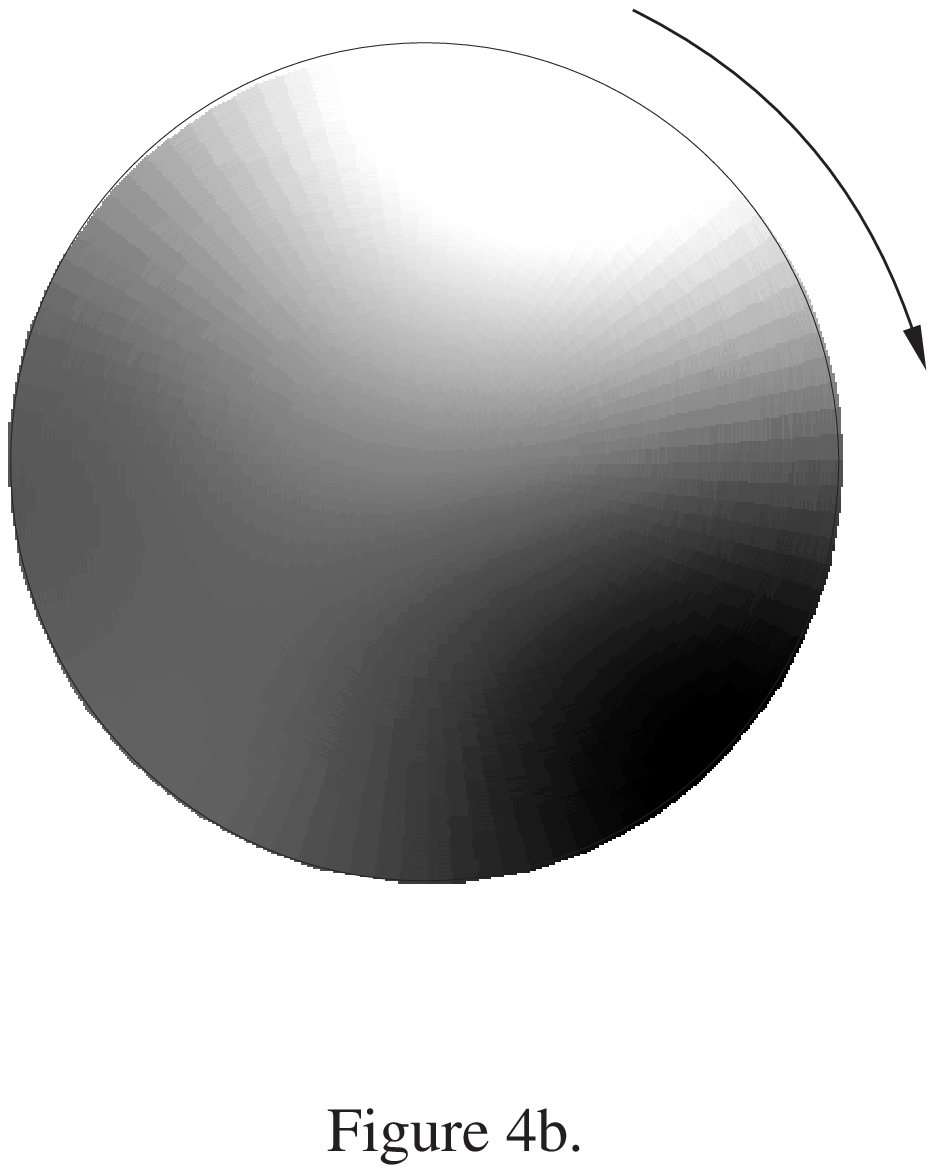}

\newpage

\epsfig{file=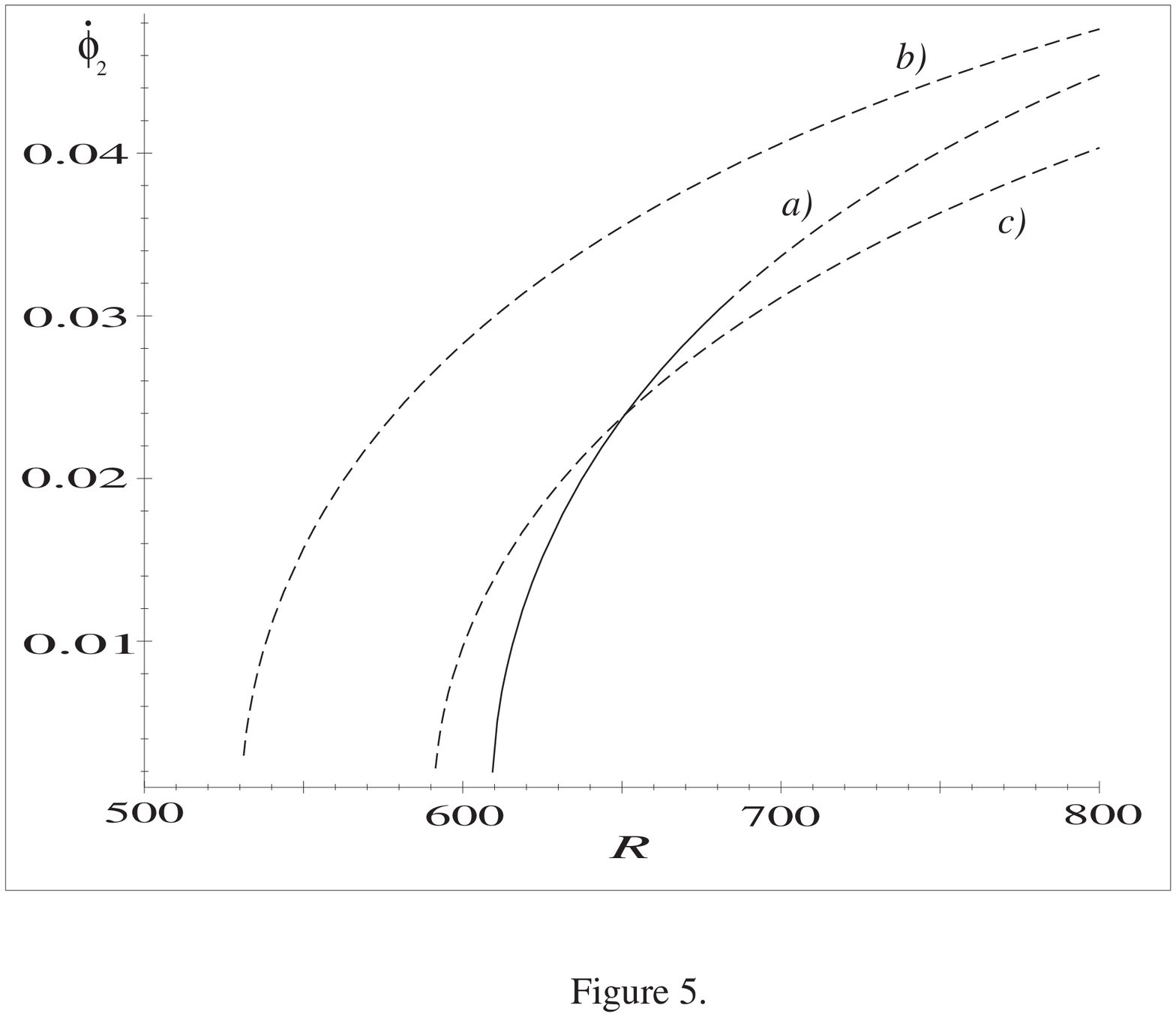,width=16cm}

\newpage

\hspace{1cm}\epsfig{file=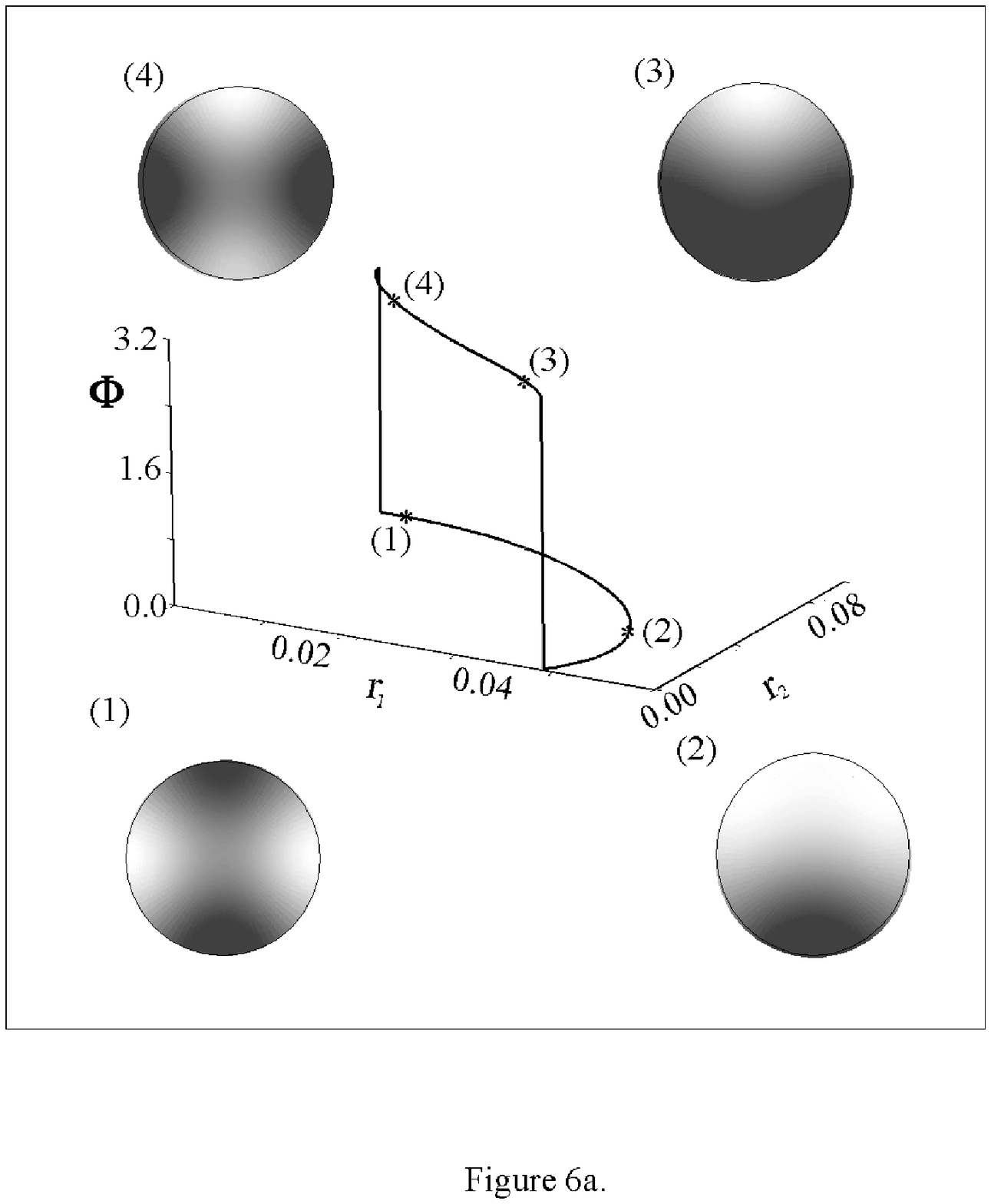}

\newpage

\epsfig{file=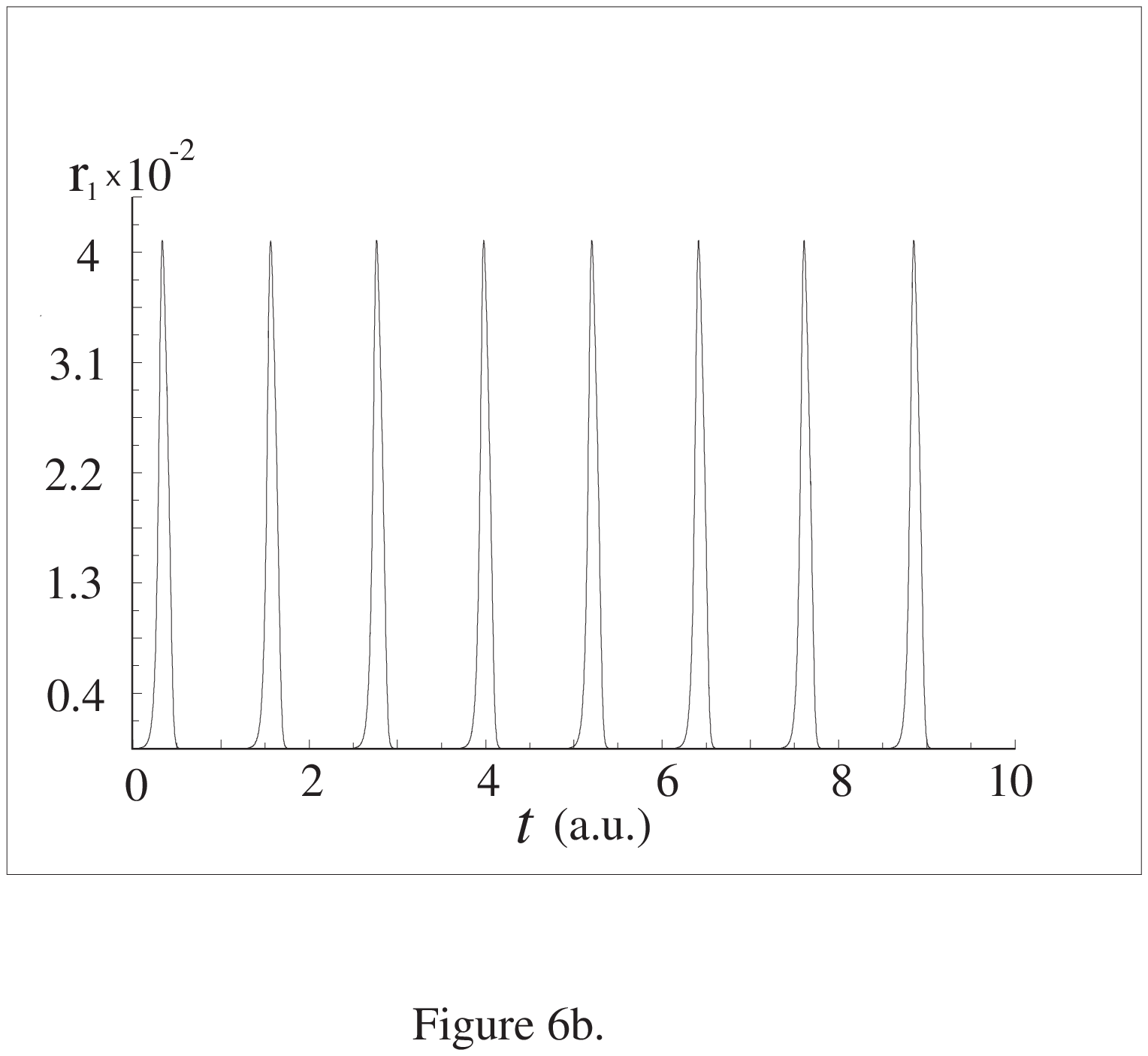}

\newpage

\epsfig{file=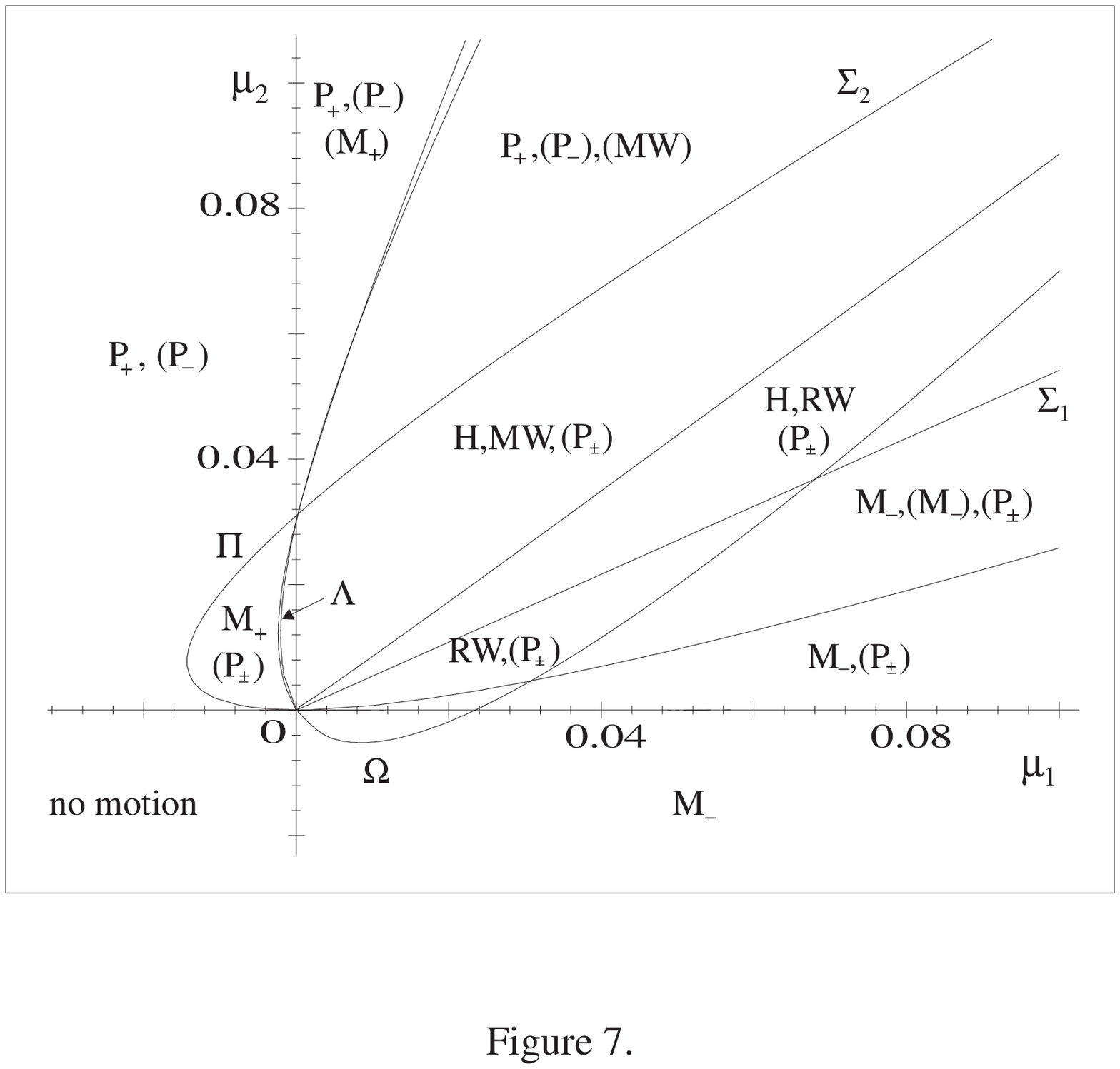}

\newpage

\epsfig{file=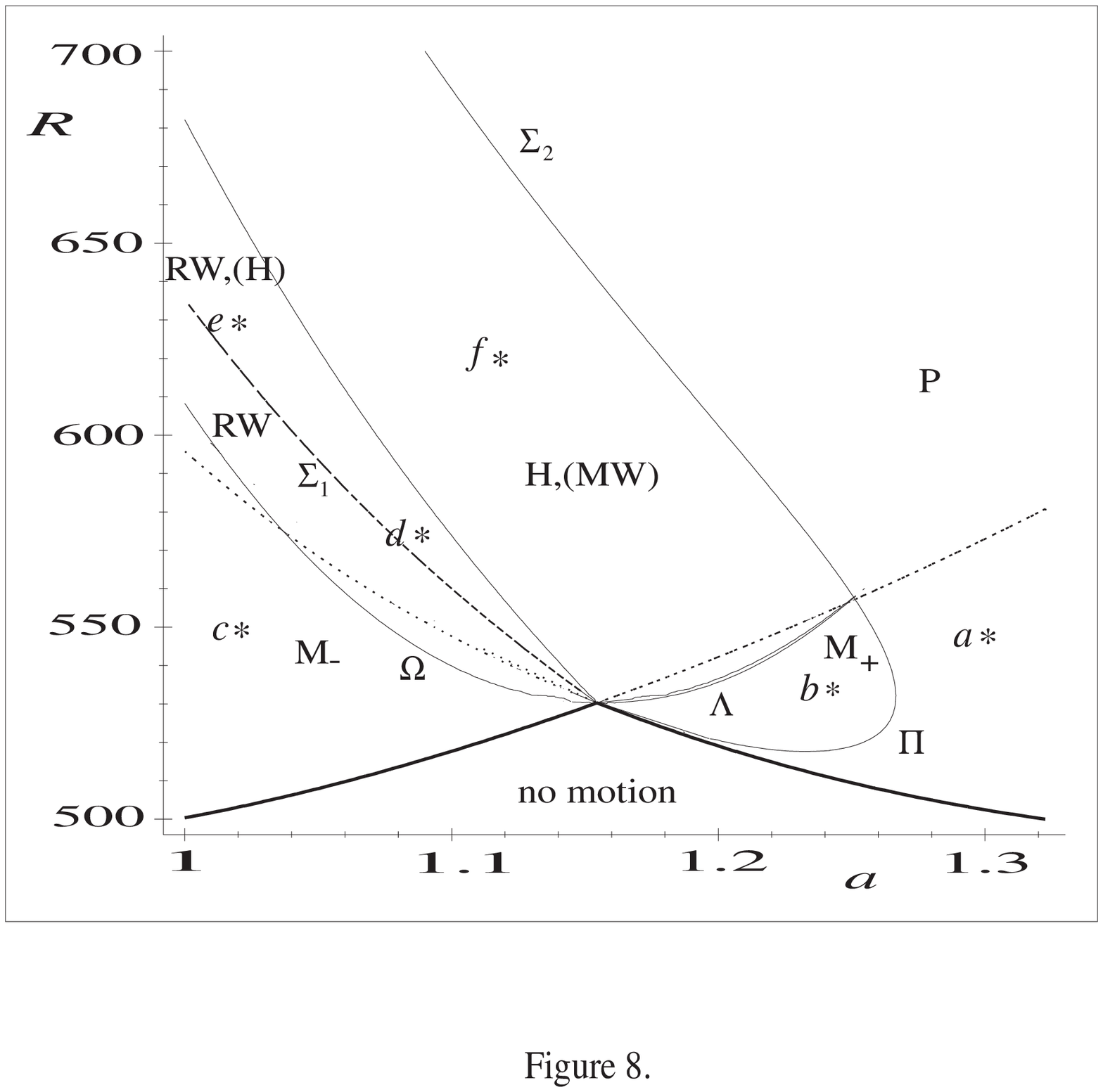}

\newpage

\epsfig{file=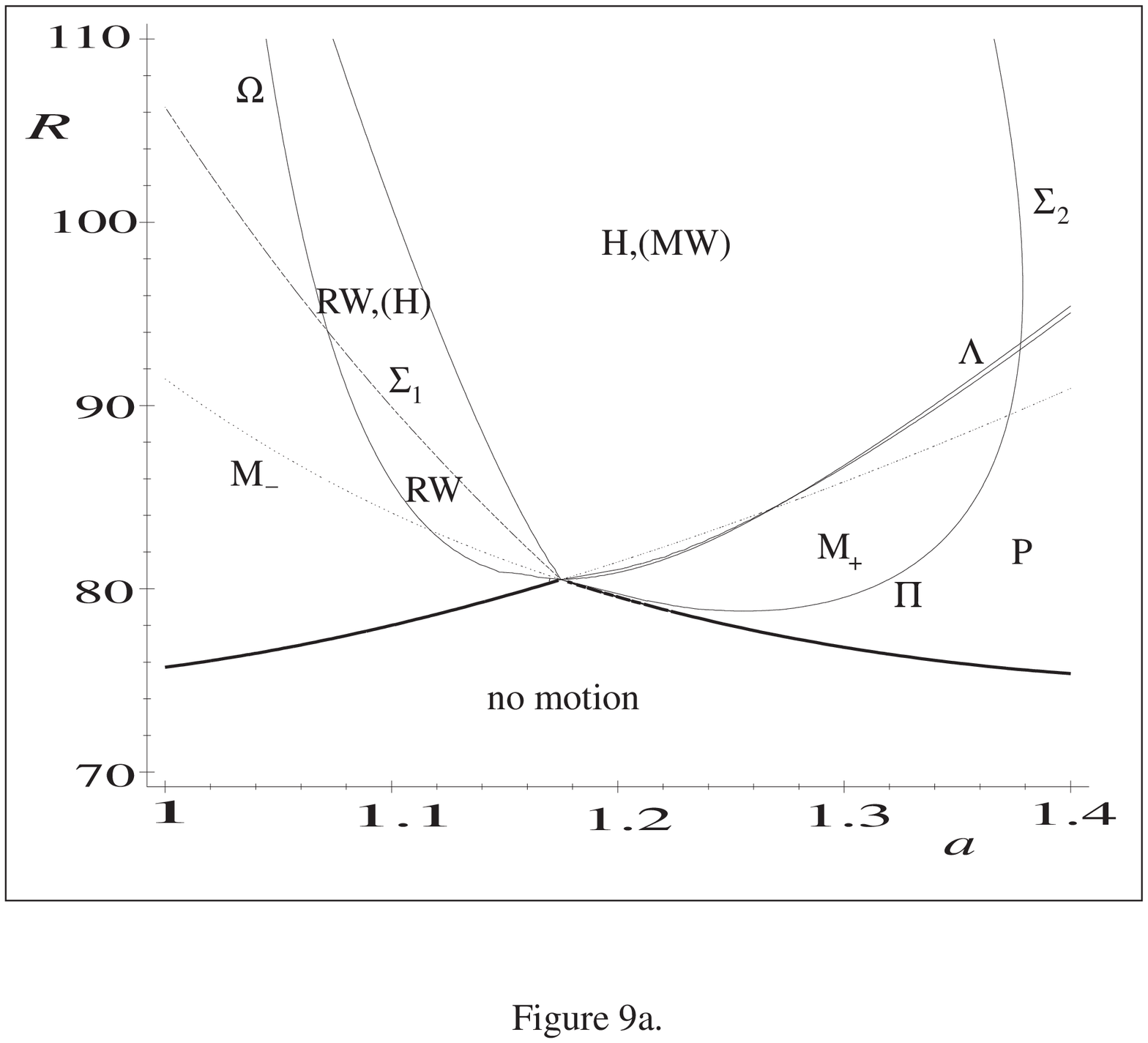,width=16cm}

\newpage

\epsfig{file=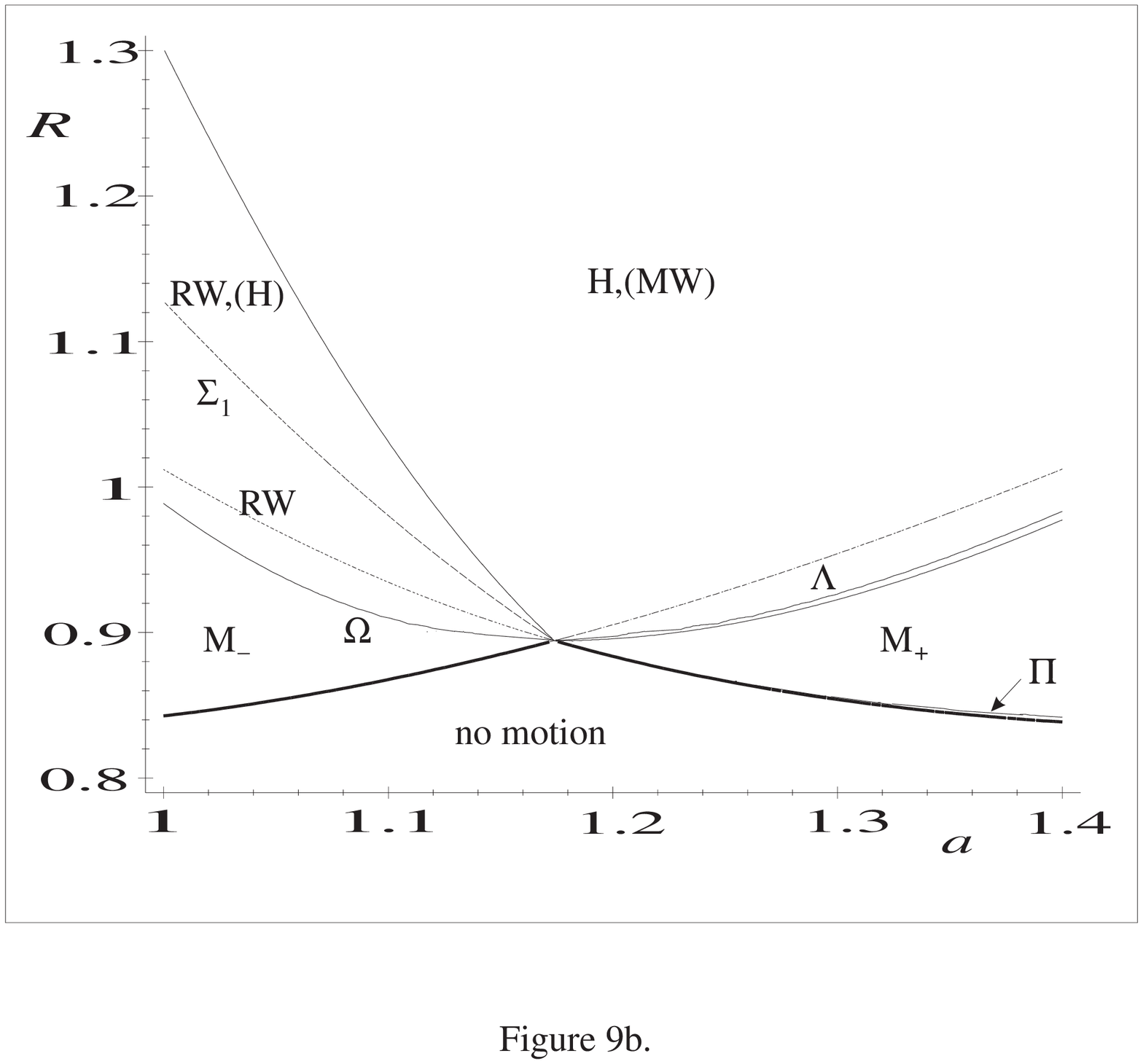,width=16cm}

\end{document}